\documentclass[letterpaper, 10 pt, conference]{ieeeconf}

\usepackage{graphicx}
\usepackage{epstopdf}
\usepackage{epsfig}
\usepackage{times}
\usepackage{amsmath}
\usepackage{amssymb}
\usepackage{latexsym}
\usepackage{color}
\usepackage{verbatim}
\usepackage{cite}
\usepackage{BernsteinStyle}
\usepackage[font = footnotesize]{caption}
\usepackage[font = footnotesize]{subcaption}
\usepackage[hidelinks]{hyperref}
\usepackage{diagbox}

\DeclareSymbolFont{matha}{OML}{txmi}{m}{it}
\DeclareMathSymbol{\varv}{\mathord}{matha}{118}

\usepackage{makecell}

\makeatletter
\newcommand*\bigcdot{\mathpalette\bigcdot@{.5}}
\newcommand*\bigcdot@[2]{\mathbin{\vcenter{\hbox{\scalebox{#2}{$\m@th#1\bullet$}}}}}
\makeatother

\IEEEoverridecommandlockouts

\title{\LARGE
Target Tracking Using the   
Invariant Extended Kalman Filter \\ with Adaptive Real-Time Numerical Differentiation for Estimating Curvature and Torsion 
}

\title{\LARGE
Target Tracking Using the   
Invariant Extended Kalman Filter \\ with  Numerical Differentiation for Estimating Curvature and Torsion 
}

\title{\LARGE
Target Tracking Using the Invariant Extended Kalman Filter \\ with Numerical Differentiation for Estimating Curvature and Torsion 
}

\author{Shashank Verma and Dennis S. Bernstein%
\thanks{$^{*}$Shashank Verma and Dennis S. Bernstein are with the Department of Aerospace Engineering, University of Michigan, Ann Arbor, MI 48109, USA 
{\tt\small shaaero@umich.edu}}%
}

\begin{document}

\maketitle
\thispagestyle{empty}
\pagestyle{empty}

\begin{abstract}
The goal of target tracking is to estimate target position, velocity, and acceleration in real time using position data.
This paper introduces a novel target-tracking technique that uses adaptive input and state estimation (AISE) for real-time numerical differentiation to estimate velocity, acceleration, and jerk from position data. 
These estimates are used to model the target motion within the Frenet-Serret (FS) frame.
By representing the model in SE(3), the position and velocity are estimated using the invariant extended Kalman filter (IEKF). 
The proposed method, called FS-IEKF-AISE, is illustrated by numerical examples and compared to prior techniques.

\end{abstract}

\section{INTRODUCTION}

The goal of target tracking is to estimate target position, velocity, and acceleration in real time.
As a key element in guidance, navigation, and control systems, target tracking relies on estimation algorithms that account for sensor noise and environmental effects.
Numerous approaches have been developed to improve the accuracy and efficiency of these systems \cite{kumar_2021_review_tt, zarchan_book_2012, BarShalom2001Estimation}.
Simplifying assumptions, such as constant velocity and constant acceleration, are often invoked. When targets exhibit complex, unpredictable maneuvers, however, more sophisticated models are needed \cite{rong_survey_tt_2003}.

Prior research has explored various approaches for improving target tracking. 
\cite{akcal_predictive_2021} employs target acceleration predictions produced by a recurrent neural network for a predictive pursuer guidance algorithm. 
\cite{breivik_straight-line_2008} discusses straight-line target tracking for unmanned surface vehicles using constant-bearing guidance to calculate a desired velocity. 
\cite{chan_kalman_1979} presents a tracking scheme based on the Kalman filter, estimating acceleration inputs from residuals and updating the filter accordingly. 
\cite{chuang_autonomous_2019} enhances real-time performance using vision-based state estimators for target tracking.

The extended Kalman filter (EKF) has been used to estimate the kinematic state of reentry ballistic targets, as shown by \cite{ghosh_tracking_2011}, while \cite{guo_maneuver_2013} uses the Kalman filter to track the movements of ballistic vehicles. 
Adaptive input estimation techniques for estimating the acceleration of maneuvering targets are demonstrated by \cite{gupta_retrospective-cost-based_2012, ahmed_input_2019, han_rcie}. 
\cite{hardiman_nonlinear_2006} compares the performance and accuracy of EKF, unscented Kalman filter, and particle filter for tracking ballistic targets and predicting impact points using measurements from 3D radar.

Iterative solutions employing state transition matrices to correct the initial conditions of ballistic vehicles for trajectory calculation have been presented by \cite{harlin_ballistic_2007}. 
To overcome the limitations of conventional constant-level maneuver models, \cite{hungu_lee_generalized_1999, shalom-1989-input-est} propose a target-tracking technique using input estimation. 
\cite{khaloozadeh_modified_2009} introduces a Kalman-filter-based tracking scheme incorporating input estimation for maneuvering targets.
\cite{lee_intelligent_2004} proposes an intelligent Kalman filter for tracking maneuvering targets, using a fuzzy system optimized by genetic algorithms and DNA coding methods to compute time$-$varying process noise. 
Machine learning techniques, such as expert prediction, have also been applied to collaborative sensor network target tracking methods by \cite{li_target_2014}.

The performance characterization of $\alpha$-$\beta$-$\gamma$ filters for constant-acceleration targets is discussed in \cite{tenne-2002-alpha-beta-gamma}, whereas \cite{Kalata1983TheTI} presents a procedure for selecting tracking parameters for all orders of tracking models.  
\cite{graymurray} provides an analytical expression for the tracking index, which is used for real-time applications. 
\cite{mookerjee_2005_reduced_state_est} presents an optimal reduced-state estimator for consistent tracking of maneuvering targets, addressing the limitations found in Kalman filters and interacting multiple model estimators commonly used for this purpose.
The accuracy of these methods depends on the $\alpha$-$\beta$-$\gamma$ parameter values. To facilitate the selection of these parameters, \cite{hasan-2013-adaptive-alpha-beta} proposes an adaptive $\alpha$-$\beta$ filter, where the $\alpha$-$\beta$ parameters are adjusted in real-time using genetic algorithms.

Recent advances in target tracking focus on improving estimation accuracy for highly maneuvering targets in 3D spaces. 
\cite{bonnabel_cdc_2017_target_tracking} explores using the Frenet-Serret frame with the invariant extended Kalman filter (IEKF) \cite{bonnabel_stable_observer_TAC_2017} to track target motion using only position measurements under assumptions of constant speed, curvature, and torsion.
\cite{marion_iekf_atc_2019} applied IEKF to air traffic control, while \cite{gibbs_fs_iekf_2022} extended the Frenet-Serret and Bishop frames within the IEKF framework to handle accelerating targets. 
However, these approaches struggle when the target trajectory, i.e., curvature, torsion, or speed, varies over time, representing realistic and complex maneuvers.
Additionally, for a state space that consists of both Lie group and vector variables, \cite{bonnabel_cdc_2017_target_tracking, pilte_iekf_tt_2019, gibbs_fs_iekf_2022}  apply  IEKF to a proper subset of the state space \cite{bonnabel_cdc_2017_target_tracking}. 
%

The contribution of this paper is a target-tracking technique that extends existing methods by removing the assumptions of constant speed, curvature, and torsion \cite{bonnabel_cdc_2017_target_tracking, pilte_iekf_tt_2019, gibbs_fs_iekf_2022}.
Unlike \cite{bonnabel_cdc_2017_target_tracking}, the state space of the proposed approach is entirely modeled in SE(3), enabling the full application of IEKF. 
Furthermore, we use the Frenet-Serret (FS) formulas to compute time$-$varying estimates of speed, curvature, and torsion. These estimates require the velocity, acceleration, and jerk of the target, which are obtained using adaptive input and state estimation (AISE) for real-time numerical differentiation.
The speed, curvature, and torsion
are used by IEKF to filter the position measurements and estimate velocity. 
Numerical examples are provided to compare the performance of the proposed method FS-IEKF-AISE with FS-IEKF developed in \cite{bonnabel_cdc_2017_target_tracking}.


This paper is organized as follows: Section \ref{sec:prob_statement} introduces the Frenet-Serret frame as target model. Section \ref{sec:iekf} discusses IEKF, while Section \ref{sec:num_example} presents three numerical examples comparing the performance of the proposed method.


\section{Problem Statement and Target Model}  \label{sec:prob_statement}
Consider a target moving in the 3D space with $\rmo_{\rm T}$ is any point fixed on the target.
We assume that the Earth is inertially non-accelerating and non-rotating.
The right-handed frame $\rmF_{\rm E} = \begin{bmatrix} \hat{\imath}_{\rmE} & \hat{\jmath}_{\rmE} & \hat{k}_{\rmE} \end{bmatrix}$ is fixed relative to the Earth, with the origin $\rmo_{\rmE}$ located at any convenient point on the Earth's surface;
hence, $\rmo_{\rmE}$ has zero inertial acceleration.
$ \hat{k}_{\rmE}$ points downward, and $\hat{\imath}_{\rmE}$ and $\hat{\jmath}_{\rmE}$ are horizontal. 
$\rmo_{\rm T}$ is any point fixed on the target.

\begin{figure}[h!t]
              \begin{center}
            {\includegraphics[width=1\linewidth]{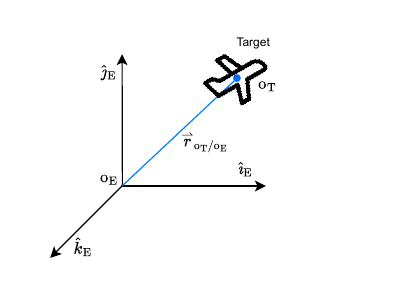}}
            \end{center}
            \caption{ $\vect{r}_{\rmo_{\rm T}/\rmo_{\rmE}}$ is the physical position vector between $\rmo_{\rm T}$ and $\rmo_\rmE$. } 
            \label{target_fig}
\end{figure}

The location of the target origin $\rmo_{\rm T}$ relative to $\rmo_{\rmE}$  is represented by the position vector $\vect{r}_{\rmo_{\rm T}/\rmo_{\rmE}}$, as shown in Figure \ref{target_fig}. 
We assume that a sensor measures the position of the target in the frame $\rmF_{\rm E}$ as
\begin{align}
    p_k \isdef \vect{r}_{\rmo_{\rm T}/\rmo_{\rmE}} (t_k) \bigg\vert_\rmE \in \mathbb{R}^3,    
     \label{pos_resolve_E}
\end{align}
where $k$ is the step and $t_k \isdef kT_\rms$. Here, $p_{k} = p(kT_{\rms})$ is the position measurements of the target moving along a trajectory $p(t) \in \mathbb{R}^3$ at step $k$, with $T_{\rms}$ being the sample time.
The first, second, and third derivatives of $\vect{r}_{\rmo_{\rm T}/\rmo_{\rmE}}$ with respect to $\rmF_{\rm E}$ represent the physical velocity, acceleration, and jerk vector $\framedotE{\vect{r}}_{\rmo_{\rm T}/\rmo_{\rmE}}$, $\frameddotE{\vect{r}}_{\rmo_{\rm T}/\rmo_{\rmE}}$, and $\framedddotE{\vect{r}}_{\rmo_{\rm T}/\rmo_{\rmE}}$. 
Resolving $\framedotE{\vect{r}}_{\rmo_{\rm T}/\rmo_{\rmE}}$, $\frameddotE{\vect{r}}_{\rmo_{\rm T}/\rmo_{\rmE}}$, and $\framedddotE{\vect{r}}_{\rmo_{\rm T}/\rmo_{\rmE}}$ in $\rmF_{\rm E}$ yields
\begin{align}
v_k \isdef \dot{p}_k = 
 \,\,\framedotE{\vect{r}}_{\rmo_{\rm T}/\rmo_{\rmE}}(t_k)\bigg\vert_\rmE \in \mathbb{R}^3, 
\label{vel_resolve_E}
\end{align}
\begin{align}
a_k \isdef \ddot{p}_k =  \, \frameddotE{\vect{r}}_{\rmo_{\rm T}/\rmo_{\rmE}}(t_k)\bigg\vert_\rmE \in \mathbb{R}^3 ,  \label{acc_resolve_E}
\end{align}
\begin{align}
j_k \isdef  \dddot{p}_k =  \, \framedddotE{\vect{r}}_{\rmo_{\rm T}/\rmo_{\rmE}}(t_k)\bigg\vert_\rmE \in \mathbb{R}^3 .  \label{jerk_resolve_E}
\end{align}

Given that the target moves along a trajectory $p(t)$, in 3D space, we model target trajectory using Frenet-Serret formulas as
\begin{align}
    \dot{p}(t) = u(t) T(t), \label{FS_pos}
\end{align}
where $T(t)$ is the unit tangent vector and $u(t) \geq 0$ is the speed along $T(t)$.
The Frenet-Serret formulas are given by
\begin{align}
    \dot{T}(t) &= u(t) \Tilde{\kappa}(t) N(t), \label{FS_eq_T} \\
    \dot{N}(t) &= u(t) [-\Tilde{\kappa}(t) T(t) + \Tilde{\tau}(t) B(t)], \label{FS_eq_N}  \\
    \dot{B}(t) &= -u(t)\Tilde{\tau}(t) N(t), \label{FS_eq_B}  
\end{align} 
where $\Tilde{\kappa}(t)$ is the curvature, $\Tilde{\tau}(t)$ is the torsion, and $N(t)$ and $B(t)$ are the unit normal and unit binormal vectors. 

The vectors $T(t)$, $N(t),$ and $B(t)$ are mutually orthogonal. For brevity, the time variable $t$ will be omitted henceforth. The Frenet-Serret formulas \eqref{FS_eq_T}, \eqref{FS_eq_N}, and \eqref{FS_eq_B},  can be written as \cite{bonnabel_cdc_2017_target_tracking}
\begin{align}
     \begin{bmatrix}
    \dot{T} & \dot{N} & \dot{B} 
    \end{bmatrix} = u \begin{bmatrix}
    T & N & B 
    \end{bmatrix} \begin{bmatrix}
    0 & -\Tilde{\kappa} & 0 \\ 
    \Tilde{\kappa} & 0 & -\Tilde{\tau}  \\ 
    0 & \Tilde{\tau} & 0
    \end{bmatrix} . \label{FS_eq_matrix}
\end{align} 
Defining $\kappa \isdef u\Tilde{\kappa}$ and $\tau \isdef u\Tilde{\tau}$, we write \eqref{FS_eq_matrix} as
\begin{align}
    \begin{bmatrix}
    \dot{T} & \dot{N} & \dot{B} 
    \end{bmatrix} =  \begin{bmatrix}
    T & N & B 
    \end{bmatrix} \begin{bmatrix}
    0 & -{\kappa} & 0 \\ 
    {\kappa} & 0 & -{\tau}  \\ 
    0 & {\tau} & 0
    \end{bmatrix} . \label{FS_eq_matrix_rewrite}
\end{align} 

Using the position, velocity, and acceleration defined by \eqref{pos_resolve_E}, \eqref{vel_resolve_E}, and \eqref{acc_resolve_E}, the tangent, normal, and binormal vectors at step $k$ are given by (excluding nongeneric cases) \cite{Hanson_FS_formula_1994} 
\begin{align}
    T_k &= \frac{p_k}{\Vert 
p_k \Vert}, \label{T_k}\\
N_k &= \frac{v_k \times (a_k \times v_k)}{\Vert 
v_k \Vert  \Vert a_k \times v_k \Vert}, \label{N_k}\\
B_k &= \frac{v_k \times a_k}{\Vert 
v_k \times a_k \Vert}, \label{B_k}
\end{align} 
where $T_k \isdef T(kT_\rms)$, $N_k \isdef N(kT_\rms),$ and $B_k \isdef B(kT_\rms).$
Similarly, $u$, $\Tilde{\kappa}$, and $\Tilde{\tau}$ at step $k$ are given by \cite{Hanson_FS_formula_1994} 
\begin{align}
    u_k &= \Vert 
v_k \Vert, \label{u_norm}\\
    \Tilde{\kappa}_k &= \frac{\Vert v_k \times a_k \Vert}{\Vert 
v_k \Vert^3}, \label{curvature_tilde}\\
\Tilde{\tau}_k &= \frac{v_k^\rmT  (a_k \times j_k)}{\Vert v_k \times a_k
\Vert^2}, \label{torsion_tilde}
\end{align}
where $u_k \isdef u(kT_\rms)$, $\Tilde{\kappa}_k \isdef \Tilde{\kappa}(kT_\rms),$  $\Tilde{\tau}_k \isdef \Tilde{\tau}(kT_\rms).$ Note
\begin{align}
    {\kappa}_k &= u_k \Tilde{\kappa}_k = u_k \frac{\Vert v_k \times a_k \Vert}{\Vert 
v_k \Vert^3}, \label{curvature}\\
    {\tau}_k &= u_k \Tilde{\tau}_k = u_k \frac{v_k^\rmT  (a_k \times j_k)}{\Vert v_k \times a_k
\Vert^2}, \label{torsion}
\end{align}
where ${\kappa}_k \isdef {\kappa}(kT_\rms)$ and ${\tau}_k \isdef {\tau}(kT_\rms)$.

We use adaptive input and state estimation (AISE) \cite{verma_shashank_2024_realtime_VRF_ACC, verma_shashank_2023_realtime_IJC} for real-time numerical differentiation to estimate derivatives of the position measurement,   
specifically, the velocity estimate $\hat{v}_k$ of \eqref{vel_resolve_E}, the acceleration estimate $\hat{a}_k$ of \eqref{acc_resolve_E}, and the jerk estimate $\hat{j}_k$ of \eqref{jerk_resolve_E}. 
Details on AISE are given in \cite{verma_shashank_2024_realtime_VRF_ACC, verma_shashank_2023_realtime_IJC}.

Using AISE to estimate $\hat{v}_k$, $\hat{a}_k$, and $\hat{j}_k$ in \eqref{u_norm}, \eqref{curvature}, and \eqref{torsion}, we obtain the Frenet-Serret parameter estimates $\hat{u}_k$, $\hat{\kappa}_k$, and $\hat{\tau}_k$ at step $k$. 
These estimates are used by the invariant extended Kalman filter (IEKF) for target tracking, as described in section \ref{sec:iekf}.

\section{Invariant Extended Kalman Filter (IEKF)} \label{sec:iekf}
In this section, we present the novel invariant extended Kalman filter (IEKF) \cite{bonnabel_stable_observer_TAC_2017, hartley_contact_aided_2020} for the tracking problem on Lie group SE(3) using only position measurement. 
To model the target in an intrinsic coordinate system based on the Frenet-Serret frame, we define the orthogonal matrix 
\begin{align}
  R \isdef \begin{bmatrix} T & N & B \end{bmatrix} \in {\rm SO}(3). \label{rot_mat}
\end{align} 
Under the assumption velocity vector $v_k$ and the tangent vector $T$ are colinear, the target dynamics using \eqref{FS_pos} and \eqref{FS_eq_matrix_rewrite} are written as
\begin{align}
    \dot{p} &= R(\nu + w_{p}),\label{pos_pred_conti} \\
     \dot{R} &= R(\omega + w_{\omega})^{\times} , \label{rot_mat_conti}
\end{align} 
where $\nu \isdef \begin{bmatrix} u & 0 & 0 \end{bmatrix}^\rmT \in \mathbb{R}^3$, $\omega \isdef \begin{bmatrix} \tau & 0 & \kappa \end{bmatrix}^\rmT = \begin{bmatrix} u\Tilde{\tau} & 0 & u\Tilde{\kappa} \end{bmatrix}^\rmT \in \mathbb{R}^3$, $(\cdot)^{\times}$ denotes a $3 \times 3$ skew-symmetric matrix, and $w_{p}, w_{\omega} \in \mathbb{R}^3$ are the white process noise. Note that the process noise for the position is only added in the tangential direction, that is, $w_p = \begin{bmatrix}
    w_{p,T} & 0 & 0
\end{bmatrix}^\rmT$, where $w_{p,T} \in \mathbb{R}$ is white noise \cite{pilte_iekf_tt_2019, gibbs_fs_iekf_2022}. 

Using the Lie group $SE(3)$, which represents the group of 3D rotations and translations, we define
\begin{align}
    \mathcal{X} \isdef \begin{bmatrix}
        R & p \\ 0_{1\times 3} & 1
    \end{bmatrix} \in SE(3). \label{se3_state}
\end{align}
Rewriting the \eqref{pos_pred_conti} and \eqref{rot_mat_conti}  using the Lie group framework \cite{gibbs_fs_iekf_2022, bonnabel_cdc_2017_target_tracking, pilte_iekf_tt_2019}
\begin{align}
    \dot{\mathcal{X}} = \mathcal{X} (\zeta + \mathcal{L}_{\mathfrak{se}(3)}(w_{\mathcal{X}})), \label{lie_group_model}
\end{align}
where, for the Lie algebra $\mathfrak{se}(3)$,
$\mathcal{L}_{\mathfrak{se}(3)}\colon\mathbb{R}^6 \rightarrow \mathfrak{se}(3)$, $w_{\mathcal{X}} = \begin{bmatrix}
    w_{\omega}^\rmT & w_{p,T} & 0 & 0
\end{bmatrix}^\rmT \in \mathbb{R}^6$, and 
\begin{align}
    \zeta \isdef \mathcal{L}_{\mathfrak{se}(3)}\bigg(\begin{bmatrix}\omega \\ \nu \end{bmatrix}\bigg) = \begin{bmatrix}
        0 & -\kappa & 0 & u \\
        \kappa & 0 & -\tau & 0 \\
        0 & \tau & 0 & 0 \\
        0 & 0 & 0 & 0
    \end{bmatrix} \in \mathfrak{se}(3).
\end{align}
The noisy position measurement $Y_k$ at step $k$ from a sensor is
\begin{align}
     Y_k = p_k + w_{Y,k}, \label{position_measurment}
\end{align} 
where $w_{Y,k} \in \mathbb{R}^3$ is   Gaussian sensor noise with  variance $S_k$. 

%
IEKF \cite{bonnabel_stable_observer_TAC_2017, gibbs_fs_iekf_2022,hartley_contact_aided_2020,bonnabel_cdc_2017_target_tracking} is applied to the target model \eqref{lie_group_model}. The evolution of the $\mathcal{X}$ in \eqref{lie_group_model} follows a symmetrical property 
\begin{align}
\dot{\mathcal{X}} = f_{\gamma}(\mathcal{X}) + \mathcal{X}w, \label{state_space_gen_X}
\end{align}
with $f_{\gamma}(\mathcal{X}) = \mathcal{X} \zeta$ satisfying 
\begin{align}
     f_{\gamma}(ab) = a f_{\gamma}(b) + f_{\gamma}(a) b  - af_{\gamma}(Id)b,
\end{align}
for all $(\gamma, a, b) \in \mathbb{R}^3 \times \rmS\rmE(3) \times \rmS\rmE(3)$, where $\gamma \isdef \begin{bmatrix} \kappa & \tau & u \end{bmatrix}$ is an input variable and $Id$ is the identity element in $\rmS\rmE(3)$. This shows that \eqref{state_space_gen_X} represents a group-affine system. Additional details are given in \cite{bonnabel_stable_observer_TAC_2017, bonnabel_cdc_2017_target_tracking}.

The system \eqref{pos_pred_conti} and \eqref{rot_mat_conti} can be written in the form of \eqref{state_space_gen_X} with
\begin{align}
    f_{\omega, \nu}\colon \begin{bmatrix}
        R & p \\
        0_{1\times 3} & 1
    \end{bmatrix} \rightarrow \begin{bmatrix}
        R\omega^\times & R\nu \\
        0_{1\times 3} & 0
    \end{bmatrix}, \label{f_map_state_space}
\end{align}
\begin{align}
    w =  \begin{bmatrix}
        w_{\omega}^{\times} & w_p \\ 0_{1\times 3} & 0
    \end{bmatrix}, \label{se3_noise}
\end{align}
and the measurement $Y_k$ \eqref{position_measurment} is re$-$written as
\begin{align}
    Y_k = \mathcal{X}_k \begin{bmatrix}
        0_{3\times1} \\ 1
    \end{bmatrix} + \begin{bmatrix}
        w_{Y,k} \\ 0
    \end{bmatrix}.
\end{align}
We treat $\omega$ and $\nu$ as external inputs to the system in \eqref{f_map_state_space}, as shown in the block diagram of FS-IEKF-AISE in Figure \ref{fig:block_diagram_FS_AISE_IEKF} below.  

Similar to the conventional Kalman filter, IEKF consists of two steps, namely, the forecast step and the data assimilation step.
In discrete time, the forecast step,
 assuming the zero-order hold between $t_k$ and $t_{k+1}$ is \cite{hartley_contact_aided_2020}
\begin{align}
    {R}_{\rmd \rma,k} &= {\mathcal{X}}_{\rmd \rma,k,[1:3,1:3]}, \label{R_da_eq}\\   
    {p}_{\rmd \rma,k} &= {\mathcal{X}}_{\rmd \rma,k,[1:3,4]},\\     
    {R}_{\rmf \rmc, k+1} &= {R}_{\rmd \rma,k} \Gamma_0(\hat{\omega}_k T_\rms), \label{R_update_k} \\
     {p}_{\rmf \rmc,k+1} &= {p}_{\rmd \rma,k} + T_\rms {R}_{\rmd \rma,k} \Gamma_1(\hat{\omega}_k T_\rms) \hat{\nu}_k, \label{pos_update_k} \\
     {\mathcal{X}}_{\rmf \rmc, k+1} &= \begin{bmatrix}
        {R}_{\rmf \rmc, k+1} & {p}_{\rmf \rmc, k+1} \\ 0_{1\times 3} & 1
    \end{bmatrix},
\end{align}
%
where ${\mathcal{X}}_{\rmd \rma,k}$, ${R}_{\rmd \rma,k}$, and ${p}_{\rmd \rma,k}$ are the data-assimilation states, ${\mathcal{X}}_{\rmf \rmc, k+1}$, ${R}_{\rmf \rmc, k+1}$, and ${p}_{\rmf \rmc, k+1}$ are the forecast states, 
$\hat{\omega}_k$, $\hat{\nu}_k$ are the estimates of $\omega_k \isdef \omega(kT\rms)$ and $\nu \isdef \nu(kT\rms)$, computed using AISE at  step $k$, and
\begin{align}
     \Gamma_{0}(\phi) \isdef I_{3} + \frac{\sin(\Vert\phi\Vert)}{\Vert\phi\Vert}\phi^{\times} + \frac{1 - \cos(\Vert\phi\Vert)}{\Vert\phi\Vert^2}\phi^{\times 2}, \label{gamma_0}
\end{align}
\begin{align}
     \Gamma_{1}(\phi) \isdef I_{3} + \frac{1 - \cos(\Vert\phi\Vert)}{\Vert\phi\Vert^2}\phi^{\times} +   \frac{\Vert\phi\Vert - \sin(\Vert\phi\Vert)}{\Vert\phi\Vert^3}\phi^{\times 2}.  \label{gamma_1}
\end{align}
%

The discrete$-$time forecast error covariance equation is given by \cite{hartley_contact_aided_2020}
\begin{align}
    {P}_{\rmf \rmc, k+1} = \Phi_{k} P_{\rmd \rma, k} \Phi_{k}^\rmT + \Tilde{Q}_{k}, \label{err_covar_pred}
\end{align}
where $P_{\rmd \rma, k}$ is the data-assimilation error covariance,  ${P}_{\rmf \rmc, k}$ is the forecast error covariance,  $\Tilde{Q}_{k} \approx \Phi_{k} Q \Phi_{k}^\rmT T_\rms$ with $Q$ is the covariance of the process noise $w = \begin{bmatrix}
    w_{\omega}^\rmT & w_{p}^\rmT
\end{bmatrix}$, and 
\begin{align}
    \Phi_{k} = {\rm exp}_\rmm(A_{k} T_\rms), \quad A_k = -\begin{bmatrix}
        \omega_k^\times & 0_{3\times 3} \\
        \nu_k^\times & \omega_k^\times
    \end{bmatrix},
\end{align}
and ${\rm exp}_\rmm(\cdot)$ is the matrix exponential \cite{gibbs_fs_iekf_2022}.

Once the measurement $Y_{k+1}$ is obtained, the data-assimilation step for IEKF is given by \cite{pilte_iekf_tt_2019, bonnabel_cdc_2017_target_tracking, gibbs_fs_iekf_2022}  
\begin{align}
    {\mathcal{X}}_{\rmd \rma, k+1} = {\mathcal{X}}_{\rmf \rmc, k+1} {\rm exp}_\rmm[\mathcal{L}_{\mathfrak{se}(3)} (K_{k+1} z_{k+1})], \label{state_update}
\end{align}
where the innovation $z_{k+1}$ is updated by
\begin{align}
    z_{k+1} = {R}_{\rmf \rmc, k+1} (Y_{k+1} - {p}_{\rmf \rmc, k+1}), \label{innov}
\end{align}
and the Kalman gain $K_{k+1}$ is
\begin{align}
    K_{k+1} = {P}_{\rmf \rmc, k+1} H^\rmT (H {P}_{\rmf \rmc, k+1} H^\rmT + {R}_{\rmf \rmc, k+1}^\rmT S_k {R}_{\rmf \rmc, k+1})^{-1
    }, 
\end{align}
where $H = \begin{bmatrix}
    0_{3\times 3} & I_{3\times 3}
\end{bmatrix}$.
The updated error covariance is 
\begin{align}
    P_{\rmd \rma, k+1} = (I_{6\times 6} - K_{k+1} H) {P}_{\rmf \rmc, k+1}. \label{err_cov_update}
\end{align}
The Kalman gain and error covariance update are derived in \cite{bonnabel_cdc_2017_target_tracking}. 
Using \eqref{FS_pos}, \eqref{rot_mat} and \eqref{R_da_eq}, the velocity estimate $\hat{v}_{\rmd \rma, k}$ is written as
\begin{align}
    \hat{v}_{\rmd \rma, k} = \hat{u}_k T_{\rmd \rma, k}.
\end{align}

%
In summary, Figure \ref{fig:block_diagram_FS_AISE_IEKF} presents a block diagram of FS-IEKF-AISE.
In FS-IEKF-AISE, we apply  IEKF to the $\rmS\rmE(3)$ system \eqref{lie_group_model}, where the Frenet-Serret parameters $\omega_k$ and $\nu_k$ are computed using AISE and used as inputs to  IEKF.

\textbf{Remark:} 
This approach differs from those in \cite{bonnabel_cdc_2017_target_tracking, pilte_iekf_tt_2019, gibbs_fs_iekf_2022} by employing AISE based numerical differentiation to estimate the time$-$varying Frenet-Serret parameters. In contrast, previous methods \cite{bonnabel_cdc_2017_target_tracking, pilte_iekf_tt_2019, gibbs_fs_iekf_2022} assume constant Frenet-Serret parameters, modeling them as Gaussian white noise.
FS-IEKF-AISE  tracks the Frenet-Serret parameters in real-time.

For comparison, the approach of \cite{bonnabel_cdc_2017_target_tracking, pilte_iekf_tt_2019} is referred to as FS-IEKF. 
In Section \ref{sec:num_example}, we compare the performance of FS-IEKF-AISE with FS-IEKF through numerical examples.

\begin{figure}[h!t]
              \begin{center}
            {\includegraphics[width=1.1\linewidth]{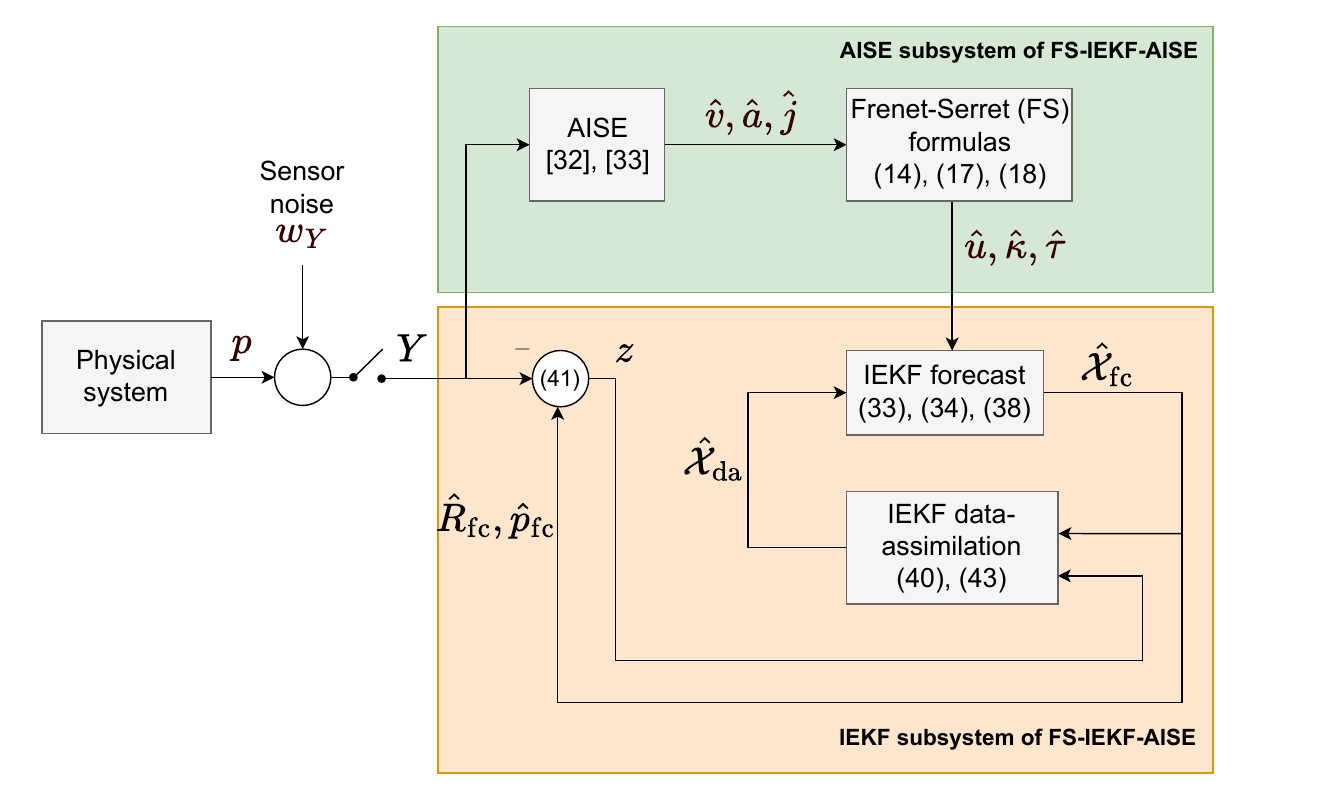}}
            \end{center}
            \caption{Block diagram of FS-IEKF-AISE} 
            \label{fig:block_diagram_FS_AISE_IEKF}
          \end{figure}

%
%

\section{Numerical Examples} \label{sec:num_example}
In this section, three numerical examples considering parabolic, helical, and Viviani (a figure$-$eight$-$shaped space curve) trajectories are provided to compare FS-IEKF and FS-IEKF-AISE. These three trajectory curves are  chosen to highlight different degree of complexity in estimating the Frenet-Serret frame parameters for the proposed methods. 
We will compare performance of target tracking methods based on the root-mean-square error (RMSE) of estimated position $\hat{p}_k$, velocity $\hat{v}_k$, curvature $\widehat{\Tilde{\kappa}}_k$, torsion $\widehat{\Tilde{\tau}}_k$, and speed $\hat{u}_k$. 

To quantify the accuracy of the target tracking methods, we define the error metric at  step $k$ as
\begin{align}
 {\rm RMSE}_{q} \isdef
\frac{1}{{N - k_0 + 1}}\sqrt{{\sum_{k=k_0}^{N}({q}_{k}-\hat{q}_{k}})^2},  \label{rms}
\end{align}
where $q$ represents the position $p = \begin{bmatrix}
    p_x & p_y & p_z
\end{bmatrix}^T$, velocity $v = \begin{bmatrix}
    v_x & v_y & v_z
\end{bmatrix}^T$, curvature $\Tilde{\kappa}$, torsion $\Tilde{\tau}$, or speed ${u}$. To account for the adaptation phase of AISE, $k$ starts from $k_0 = 1000$ in \eqref{rms}.

\begin{exam} \label{eg:traj_extra_parabola}
      {\it Target Tracking for a Parabolic Trajectory.}
      {\rm In this simulation scenario, the target follows a parabolic trajectory in the planar model, i.e., we only consider $x$ and $y$ axis, under constant gravity of $9.8$ m/s$^2$ in the negative $y$ direction. Note that the simulation accounts for the $z$ axis, representing a 3D scenario. However, since the position along the $z$ axis remains at zero, we  consider only the $x$-$y$ planar model. The discrete$-$time equations governing the trajectory are defined as 
      \begin{align}
      p_k = \begin{bmatrix}
           400 k T_{\rms} \\ 400 k T_{\rms} - 9.8 \frac{(kT_\rms)^2}{2} \\ 0
      \end{bmatrix}, \label{parabolic_traj}
    \end{align}
    where $T_{\rms} = 0.01$ s.  To simulate the noisy measurement of the target position $p_{k}$, white Gaussian noise is added to position measurement, with standard deviation $\sigma = 1.0$ m, which,  for a meaningful comparison, is of the same order as in \cite{gibbs_fs_iekf_2022,bonnabel_cdc_2017_target_tracking}.

    For  FS-IEKF  \cite{bonnabel_cdc_2017_target_tracking}, $\mathcal{X}_0 = I_{4\times4} \in {\rm SE(3)}$, $\mathcal{Z}_0 \in \mathbb{R}^3$ is set to a value that is within a small margin of the true value,
    %
    $P_0 = I_{9\times9}$, $Q = 10^{-3} {\rm diag}(0.2, 0.2, 0.2, 1, 1, 0.01, 1, 1, 1)$, and $S = {\rm diag}(1, 1, 10^{-8})$. 
    For FS-IEKF-AISE with single differentiation,  \cite{verma_shashank_2023_realtime_IJC, verma_shashank_2024_realtime_VRF_ACC}, we set $n_\rme = 25$, $n_\rmf = 50$, $R_z = 1$, $R_d = 10^{-1}$, $R_\theta = 10^{-3.5}I_{51}$, $\eta = 0.002, \tau_n = 5, \tau_d = 25, \alpha = 0.002$, and $R_{\infty} = 10^{-4}.$ 
    The parameters $V_{1,k}$ and $V_{2,k}$ are adapted, with $\eta_{L} = 10^{-6}$, $\eta_{\rmU} = 0.1$, and $\beta = 0.55$. 
    For double differentiation using AISE, the parameters are the same as those used for single differentiation. For triple differentiation, the parameters for AISE are the same as those used for single differentiation except that $R_\theta = 10^{-6}I_{51}$ and $\beta = 0.48$. 
    The initial state $\mathcal{X}_0 = I_{4\times4}$ and $P_0 = I_{6\times6}$. 
    %
    To ensure a meaningful comparison, the process-noise and measurement-noise covariances are the same as those used by FS-IEKF.


    The derivative estimated by AISE are filtered with a 4$^{th}$-order lowpass Butterworth filter with a cutoff frequency of $10$ Hz. 
    This filter is used in all examples considered, irrespective of the noise level and signal type. 

    Table \ref{table:rmse_values_parabolic} shows the RMSE \eqref{rms} for position, velocity, and the Frenet-Serret frame parameters for FS-IEKF and FS-IEKF-AISE. 
    As shown, the RMSE for FS-IEKF-AISE is consistently lower across all metrics (position, velocity, and Frenet Serret frame parameters) when compared to FS-IEKF.
    Figure \ref{fig:exp_parabola_AISE_FS_2d} shows the estimated position, with (b) providing a zoomed view of (a), showing that FS-IEKF-AISE closely tracks the true position compared to FS-IEKF. Figure \ref{fig:exp_parabola_AISE_FS_parameter} shows the estimated Frenet-Serret parameters.

\begin{table}[h!t]
\begin{center} 
\begin{tabular}{|c|c|c|c|c|c|c|}
\hline
\textbf{\backslashbox{RMSE}{Method}} & \textbf{FS-IEKF} & \textbf{FS-IEKF-AISE}  \\
\hline
 {${\rm RMSE}_{p_x}$} & 10.95 & {2.75}  \\
\hline
{${\rm RMSE}_{p_y}$}& 7.24 &  {1.77} \\
\hline
 {${\rm RMSE}_{v_x}$} & 9.17 & 1.96  \\
\hline
{${\rm RMSE}_{v_y}$}& 6.82 &  1.67 \\
\hline
{${\rm RMSE}_{\Tilde{\kappa}}$} & {5.5e$-$5} &  {2.5e$-$5} \\
\hline
{${\rm RMSE}_{\Tilde{\tau}}$}& 2.7e$-$3 &  {1e$-$10} \\
\hline
{${\rm RMSE}_{u}$}& 11.30 & {1.95}  \\
\hline
\end{tabular}
\end{center} 
\caption{{\it  Example \ref{eg:traj_extra_parabola}: Target tracking for a parabolic trajectory.} ${\rm RMSE}_{q}$ \eqref{rms} for parabolic trajectory.}
\label{table:rmse_values_parabolic}
\end{table}

 \begin{figure}[h!t]
              \begin{center}{\includegraphics[width=1\linewidth]{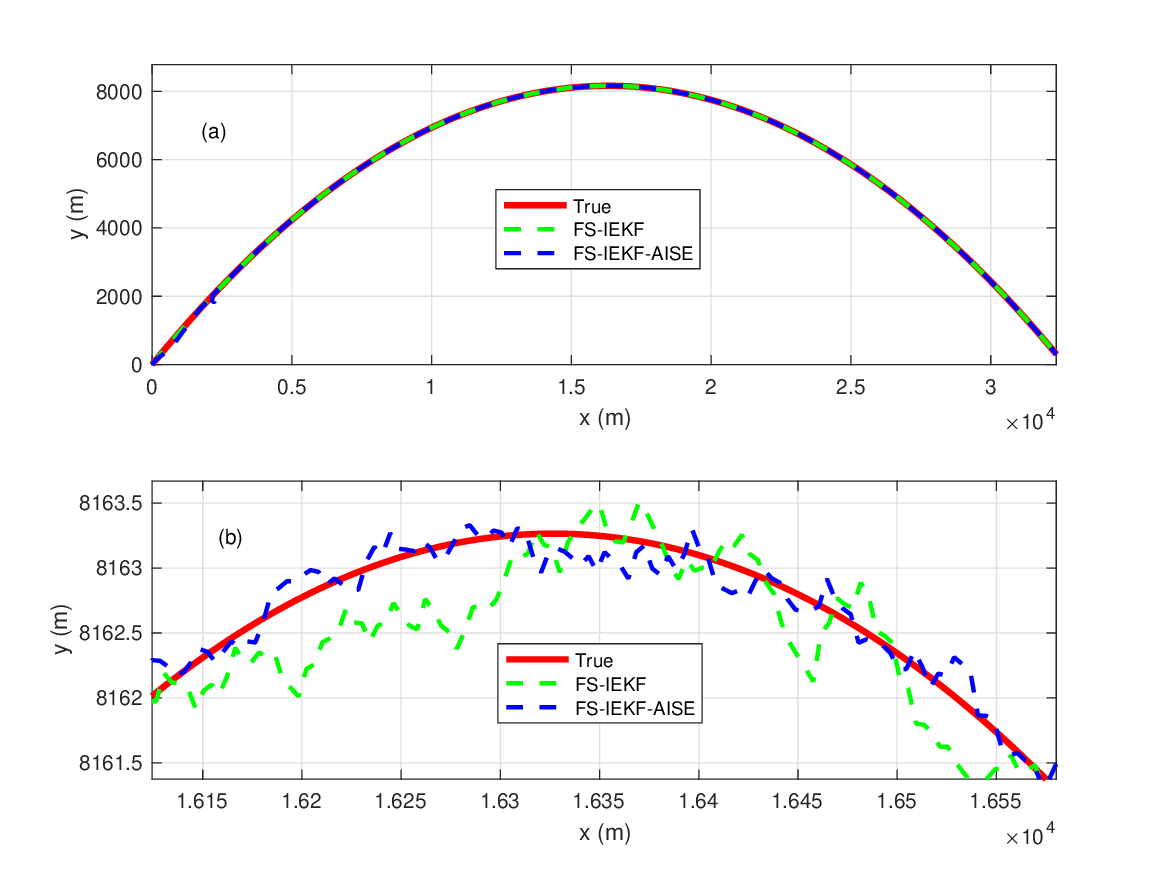}}
            \end{center}
            \caption{ {\it  Example \ref{eg:traj_extra_parabola}: Target tracking for a parabolic trajectory.} (a) Trajectory plot $x-y$.   (b) Zoom of (a). FS-IEKF-AISE in blue closely follows true trajectory in red as compare to FS-IEKF in green.} 
        \label{fig:exp_parabola_AISE_FS_2d}
          \end{figure}

 \begin{figure}[h!t]
              \begin{center}
            {\includegraphics[width=1\linewidth]{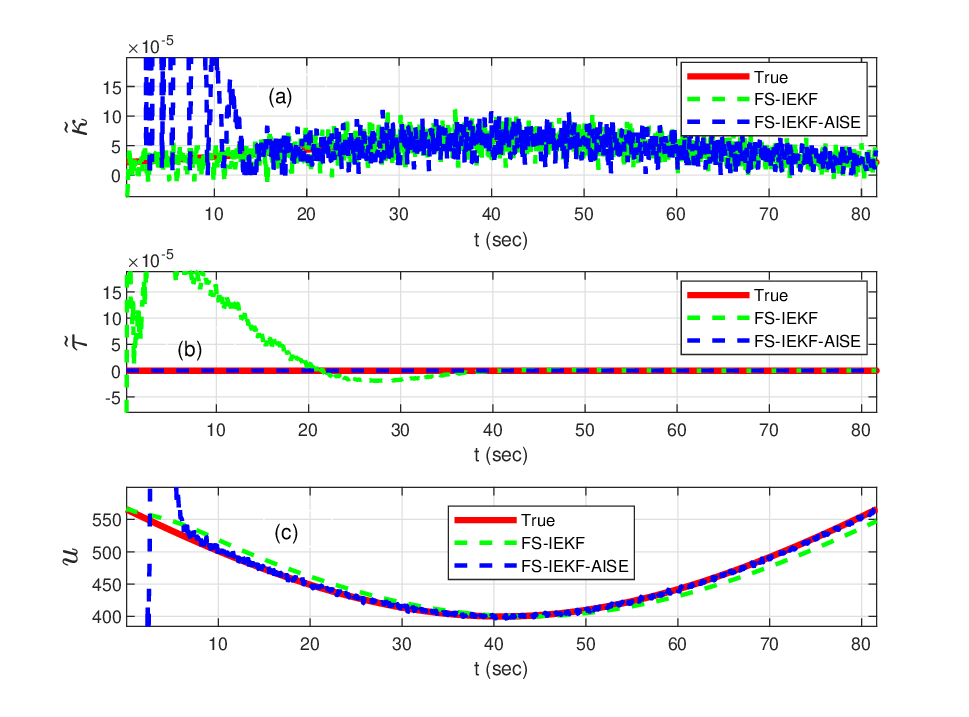}}
            \end{center}
            \caption{ {\it  Example \ref{eg:traj_extra_parabola}: Target tracking for a parabolic trajectory.} Estimates of $\Tilde{\kappa}$, $\Tilde{\tau}$, and $u$ computed using  FS-IEKF  and FS-IEKF-AISE. (a) shows the estimates of $\Tilde{\kappa}$, (b) shows the estimates of $\Tilde{\tau}$, and (c) shows the estimates of $u$. The estimates obtained using FS-IEKF-AISE are closer to the true value than those of FS-IEKF.} 
            \label{fig:exp_parabola_AISE_FS_parameter}
          \end{figure}

}      
\end{exam}

\begin{exam} \label{eg:traj_extra_helix}
      {\it Target Tracking for a Helical Trajectory.}
      {\rm In this simulation scenario, the target follows a helical trajectory. The discrete$-$time equations governing the trajectory are defined as 
      \begin{align}
      p_k = \begin{bmatrix}
           20 \sin ( k T_{\rms}) \\ 20 \cos ( k T_{\rms}) \\ k T_{\rms}
      \end{bmatrix}, \label{helix_traj}
    \end{align}
    where $T_{\rms} = 0.01$ s.  To simulate the noisy measurement of the target position $p_{k}$, white Gaussian noise is added to position measurement, with standard deviation $\sigma = 0.5$ m.

    For  FS-IEKF,  $\mathcal{X}_0 = I_{4\times4}$,  $\mathcal{Z}_0$ is set to a value that is within a small margin of the true value, $P_0 = I_{9\times9}$, $Q = 10^{-3} {\rm diag}(0.2, 0.2, 0.2, 1, 1, 0.01, 1, 1, 1)$, and $S = {\rm diag}(0.1, 0.1, 0.1)$. 
    For FS-IEKF-AISE with single differentiation, the parameters are the same as those used in Example \ref{eg:traj_extra_parabola} except that $R_d = 10^{-7}$. 
    For the cases of double and triple differentiation using AISE and IEKF, the parameters are the same as those used in Example \ref{eg:traj_extra_parabola}.

    Table \ref{table:rmse_values_helix} shows the RMSE values \eqref{rms} for the position, velocity, and Frenet-Serret frame parameters for both FS-IEKF and FS-IEKF-AISE.  FS-IEKF-AISE achieves performance comparable to FS-IEKF, as expected. This is because, for the helical trajectory, the Frenet-Serret parameters remain constant over time, making the constant parameter assumption in FS-IEKF valid for this case. Figure \ref{fig:exp_helix_AISE_FS_3d} shows the estimated position. Figure \ref{fig:exp_helical_AISE_FS_parameter} shows the estimated Frenet-Serret parameters using FS-IEKF and FS-IEKF-AISE. 
\begin{table}[h!t]
\begin{center} 
\begin{tabular}{|c|c|c|c|c|c|c|c|c|}
\hline
\textbf{\backslashbox{RMSE}{Method}} & \textbf{FS-IEKF} & \textbf{FS-IEKF-AISE} \\
\hline
 {${\rm RMSE}_{p_x}$} & 0.111 & 0.105 \\
\hline
{${\rm RMSE}_{p_y}$} & 0.107 &  0.111 \\
\hline
{${\rm RMSE}_{p_z}$} & 0.121 &  0.078 \\
\hline
 {${\rm RMSE}_{v_x}$} & 0.309 & 0.588 \\
\hline
{${\rm RMSE}_{v_y}$} & 0.292 & 0.536 \\
\hline
{${\rm RMSE}_{v_z}$} & 0.401 &  0.431 \\
\hline
{${\rm RMSE}_{\Tilde{\kappa}}$} & 0.0012 &  0.0049 \\
\hline
{${\rm RMSE}_{\Tilde{\tau}}$}& {0.001} & 0.003 \\
\hline
{${\rm RMSE}_{u}$}  & 0.232  & 0.274 \\
\hline
\end{tabular}
\end{center} 
\caption{{\it  Example \ref{eg:traj_extra_helix}: Target tracking for a helical trajectory.} ${\rm RMSE}_{q}$ \eqref{rms} for helical trajectory.}
\label{table:rmse_values_helix}
\end{table}

 \begin{figure}[h!t]
              \begin{center}
            {\includegraphics[width=0.8\linewidth]{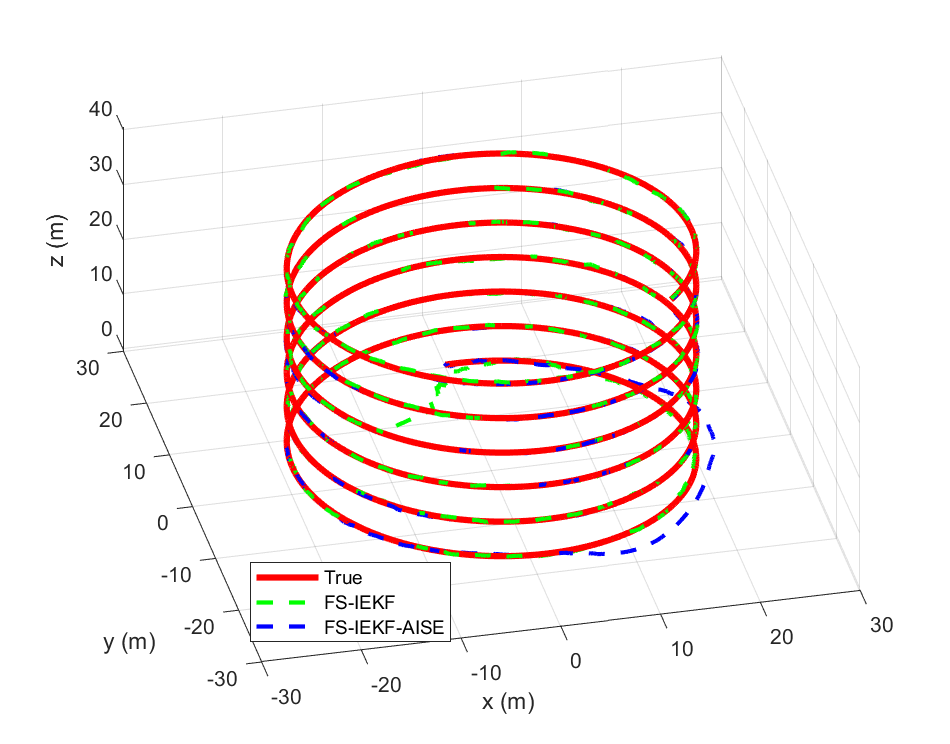}}
            \end{center}
            \caption{ {\it  Example \ref{eg:traj_extra_helix}: Target tracking for a helical trajectory.} 3D Trajectory plot. FS-IEKF-AISE in blue closely follows the true trajectory in red, similar to FS-IEKF in green.} 
            \label{fig:exp_helix_AISE_FS_3d}
          \end{figure}

 \begin{figure}[h!t]
              \begin{center}
            {\includegraphics[width=1\linewidth]{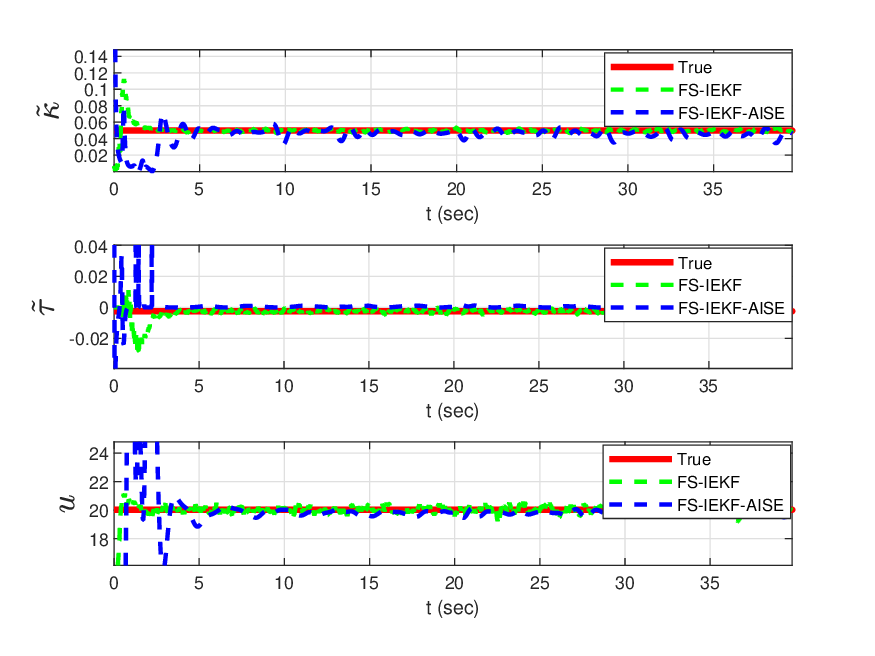}}
            \end{center}
            \caption{ {\it  Example \ref{eg:traj_extra_helix}: Target tracking for a helical trajectory.} Estimates of $\Tilde{\kappa}$, $\Tilde{\tau}$, and $u$ computed using  FS-IEKF  and FS-IEKF-AISE. (a) shows the estimates of $\Tilde{\kappa}$, (b) shows the estimates of $\Tilde{\tau}$, and (c) shows the estimates of $u$.} 
            \label{fig:exp_helical_AISE_FS_parameter}
            \vspace{-5 mm}
          \end{figure}

}      
\end{exam}

\begin{exam} \label{eg:traj_extra_viviani}
      {\it Target Tracking for the Viviani Trajectory.}
      {\rm In this simulation scenario, the target follows the Viviani trajectory. The discrete$-$time equations governing the trajectory are defined as 
      \begin{align}
      p_k = 200\begin{bmatrix}
            \cos^2 (k T_{\rms}) \\  \sin^2 (k T_{\rms}) \\  \sin(k T_{\rms})
      \end{bmatrix}, \label{viviani_traj}
    \end{align}
    where $T_{\rms} = 0.01$ s. To simulate the noisy measurement of the target position $p_{k}$, white Gaussian noise is added to position measurement, with standard deviation $\sigma = 10$ m.

    For  FS-IEKF, $\mathcal{X}_0 = I_{4\times 4}$, and $\mathcal{Z}_0$ is set to a value that is within a small margin of the true value, $P_0 = I_{9\times9}$,  $Q = 0.1 {\rm diag}(0.2, 0.2, 0.2, 1, 1, 0.01, 1, 1, 1)$, and $S = \text{diag}(10, 10, 10)$. In this example, the measurement noise is high, with a signal-to-noise ratio (SNR) of 20 dB.
    For FS-IEKF-AISE, for single differentiation using AISE, the parameters are the same as those used in Example \ref{eg:traj_extra_parabola} except that $R_d = 10^{-7}$. 
    All remaining parameters are same as in Example \ref{eg:traj_extra_parabola}.  

    Table \ref{table:rmse_values_viviani} shows the RMSE values \eqref{rms}.  FS-IEKF-AISE consistently achieves the lower RMSE values for all metrics except for ${\rm RMSE}_{\Tilde{\tau}}$ when compared to FS-IEKF.
    %
    %
    Figure \ref{fig:exp_Viviani_AISE_FS_3d} shows the estimated position. Figure \ref{fig:exp_Viviani_AISE_FS_parameter} shows that, after the initial transient, the estimated Frenet-Serret parameters provided by FS-IEKF-AISE are more accurate than the estimates of FS-IEKF.

\begin{table}[h!t]
\begin{center} 
\begin{tabular}{|c|c|c|c|c|c|c|}
\hline
\textbf{\backslashbox{RMSE}{Method}} & \textbf{FS-IEKF} & \textbf{FS-IEKF-AISE}  \\
\hline
 {${\rm RMSE}_{p_x}$} & 9.53 & 3.72  \\
\hline
{${\rm RMSE}_{p_y}$}& 9.59 & 3.94 \\
\hline
{${\rm RMSE}_{p_z}$}&  12.54 &  4.15 \\
\hline
 {${\rm RMSE}_{v_x}$} & 48.62 & 14.52  \\
\hline
{${\rm RMSE}_{v_y}$} & 45.57 & 14.15  \\
\hline
{${\rm RMSE}_{v_z}$} & 51.22 &  12.42  \\
\hline
{${\rm RMSE}_{\Tilde{\kappa}}$} & 0.015 &  0.0013 \\
\hline
{${\rm RMSE}_{\Tilde{\tau}}$}& 0.008 &  0.015 \\
\hline
{${\rm RMSE}_{u}$} & 53.323 & 10.634  \\
\hline
\end{tabular}
\end{center} 
\caption{{\it  Example \ref{eg:traj_extra_viviani}: Target tracking for the Viviani trajectory.} ${\rm RMSE}_{q}$ \eqref{rms} for the Viviani trajectory.}
\label{table:rmse_values_viviani}
\vspace{-4 mm}
\end{table}

 \begin{figure}[h!t]
              \begin{center}
            {\includegraphics[width=0.8\linewidth]{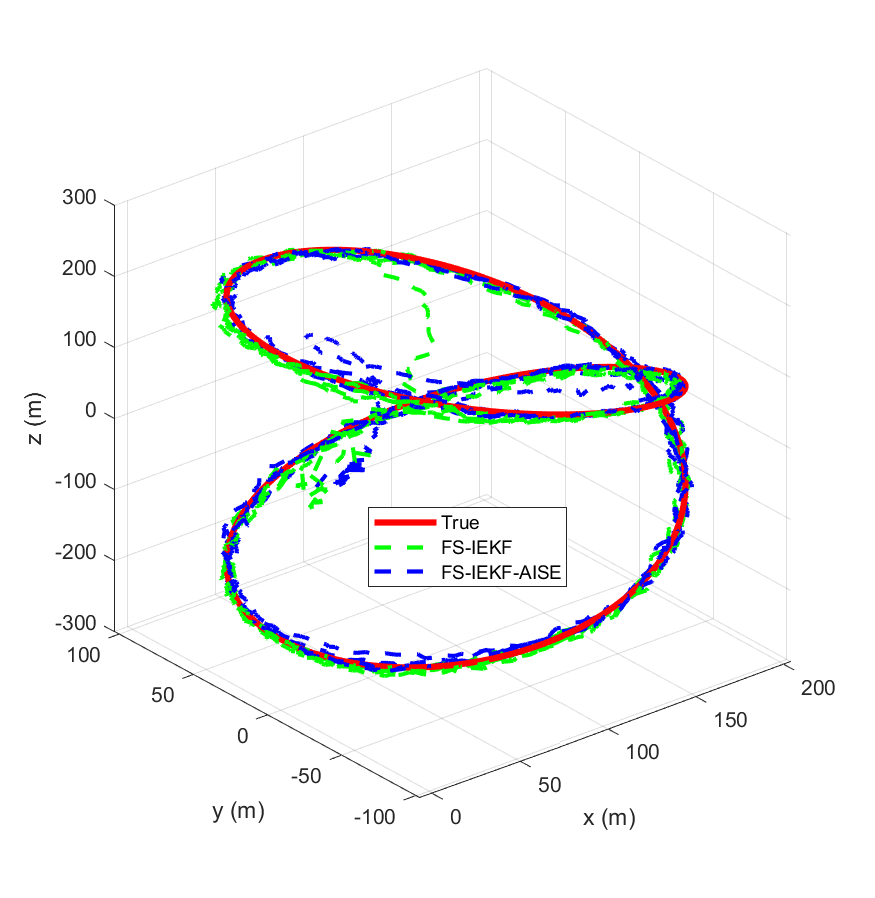}}
            \end{center}
            \caption{ {\it  Example \ref{eg:traj_extra_viviani}: Target tracking for the Viviani trajectory.} 3D trajectory plot. FS-IEKF-AISE in blue closely follows the true trajectory in red, similar to FS-IEKF in green.} 
            \label{fig:exp_Viviani_AISE_FS_3d}
            \vspace{-5 mm}
          \end{figure}

 \begin{figure}[h!t]
              \begin{center}
            {\includegraphics[width=1\linewidth]{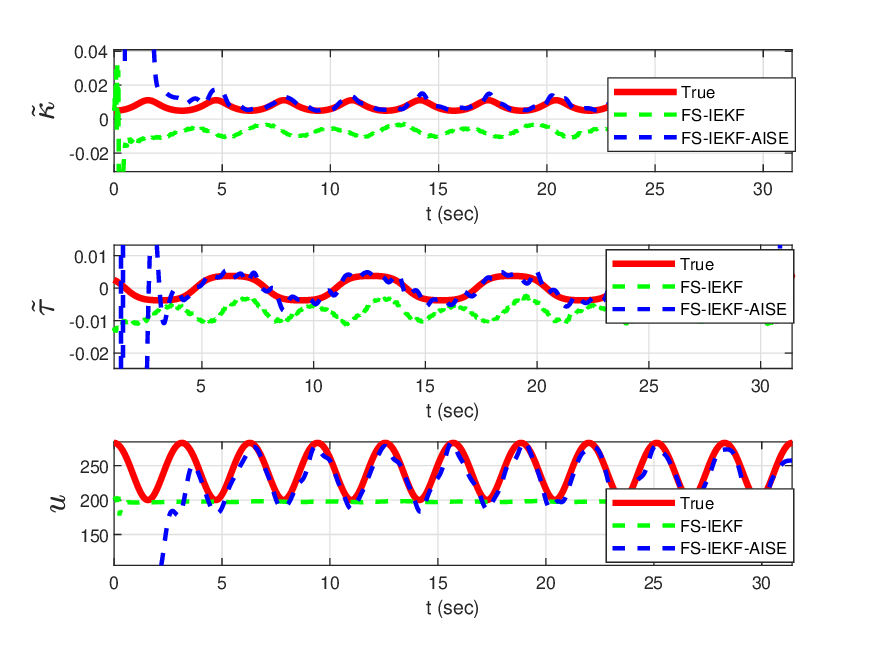}}
            \end{center}
            \caption{ {\it  Example \ref{eg:traj_extra_viviani}: Target tracking for the Viviani trajectory.} Estimates of $\Tilde{\kappa}$, $\Tilde{\tau}$, and $u$ computed using FS-IEKF and FS-IEKF-AISE. (a) shows the estimates of $\Tilde{\kappa}$, (b) shows the estimates of $\Tilde{\tau}$, and (c) shows the estimates of $u$. The estimates obtained by FS-IEKF-AISE are more accurate than those of FS-IEKF.} 
            \label{fig:exp_Viviani_AISE_FS_parameter}
            \vspace{-5 mm}
          \end{figure}

}      
\end{exam}

\section{Conclusions and Future Research}

This paper proposed a novel target-tracking technique called FS-IEKF-AISE, which integrates the Frenet-Serret (FS) frame with adaptive input and state estimation (AISE) and the Invariant Extended Kalman Filter (IEKF).  AISE  is employed for real-time numerical differentiation on noisy position data to estimate velocity, acceleration, and jerk. These estimates are crucial for dynamically defining the torsion and curvature parameters of the FS frame, enabling the system to  model complex, time$-$varying target trajectories without assuming constant motion parameters.

%
%

By applying  IEKF to the FS frame, FS-IEKF-AISE achieves better accuracy compared to previous methods in estimating the target position and velocity. This approach not only enhances tracking performance in scenarios involving highly maneuverable targets but also preserves the group-affine structure of the system model, which has empirically improved convergence. Additionally, the FS-IEKF-AISE method ensures that the state evolution remains consistent within the Lie group framework.

Future work will focus on  validating  FS-IEKF-AISE  in real-world applications. Finally, developing an adaptive IEKF that  dynamically tunes the noise covariance will further enhance the accuracy of FS-IEKF-AISE in uncertain, time$-$varying operational conditions.

\section*{ACKNOWLEDGMENTS}
This research supported by NSF grant CMMI 2031333.

\bibliography{bib_paper,bib_target_tracking}
\bibliographystyle{ieeetr}

\end{document}